\documentclass{emulateapj}

\begin{document}
\title{The Formation of Polar Disk Galaxies} 
\author{Chris B. Brook\altaffilmark{1}, Fabio Governato \altaffilmark{1}, Thomas Quinn \altaffilmark{1}, James Wadsley \altaffilmark{2}, \\Alyson M. Brooks \altaffilmark{1}, Beth Willman \altaffilmark{3}, Adrienne Stilp \altaffilmark{1}, Patrik Jonsson \altaffilmark{4}}

\altaffiltext{1}{Department of Astronomy, University of Washington, Box 351580, Seattle, WA 98195, USA}
\altaffiltext{2}{ Department of Physics and Astronomy, McMaster University, Hamilton, ON, L88 4M1, Canada}
\altaffiltext{3}{Harvard-Smithsonian Centre for Astrophysics, Cambridge, MA (Clay Fellow) }
\altaffiltext{4}{Physics Department, University of California Santa Cruz, CA 95064, USA}


\begin{abstract}

Polar Ring Galaxies, such as  NGC4650A, are a class of galaxy which have two kinematically distinct components that  are inclined by almost 90 degrees to each other. These striking galaxies challenge our understanding of how galaxies form; the origin of their distinct components has remained uncertain, and  the subject of much debate. We use high-resolution cosmological simulations of galaxy formation to show that  Polar Ring Galaxies are simply an extreme example of the angular moment misalignment that occurs during the hierarchical structure formation characteristic of Cold Dark Matter cosmology. In our model, Polar Ring Galaxies  form through the continuous accretion of gas whose angular momentum is misaligned with the central galaxy.  

\end{abstract}

\section{Introduction}
Significant numbers of galaxies are known to have  more than one kinematically distinct component, distinguishing them as  ``multi-spin" galaxies \cite{rubin}.  In some cases,  these distinct kinematic structures are essentially perpendicular to each other, and such galaxies are traditionally  referred to as Polar Ring Galaxies \cite{whitmore}. The formation of these  exotic astronomical phenomena has been the subject of much debate \cite{casertano}, and their  existence  has presented a logical puzzle to the way we believe that galaxies may form.  Interest in Polar Ring Galaxies is further heightened  by their use in probing the nature of Dark Matter \cite{schweizer,sackett,iodice06}. Although it comprises 85$\%$ of the matter in the Universe, we know little about the nature of Dark Matter. Vital clues can be found by probing its role in forming these fascinating and quixotic galaxies which have two perpendicular components.

The most extensively observed Polar Ring Galaxies share many properties with disk galaxies, including exponential light profiles (Schweizer Whitmore \& Ruben 1983), large amounts (several times $10^9 M_\odot$) of neutral Hydrogen (HI) (van Driel et al. 2000; Arnaboldi et al. 1995) in extended rather than narrow ring structures (van Gorkom Scechter \& Kristian 1987; van Driel et al. 1995; Arnaboldi et al. 1997; van Driel et al. 2000; Iodice et al. 2002; Gallagher et al. 2002) a ratio of HI mass to
luminosity in the B band (M(HI)/LB) typical of late-type spirals (Huchtmeier 1997;
Arnaboldi et al. 1997; Sparke \& Cox 2000; Cox Sparke \& van Moorsel 2006), young stellar populations (Gallagher et al. 2002; Karataeva et al. 2004a; Cox Sparke \& van Moorsel 2006), ongoing and
continuous rather than bursty star formation (Reshetnikov Faundez-Abans \& de Oliviera-Abans 2002;
Karataeva et al. 2004b), disk galaxy colors and color gradients (Reshetnikov Hagen-Thorn \& Yakovleva 1994; Arnaboldi et al. 1995), flat rotation curves (Reshetnikov Faundez-Abans \& de Oliviera-Abans 2002; Swaters \& Rubin 2003), chemical abundances typical of disk galaxies (Buttiglione Arnaboldi \& Iodice 2006), and spiral arms (Arnaboldi et al. 1995; Iodice et al. 2004; Cox Sparke \& van Moorsel 2006). So compelling has been the evidence linking polar structures to disks in galaxies such as NGC4650A that a change of nomenclature is occurring, with increasing reference to
Polar Disk Galaxies rather than Polar Ring Galaxies (Iodice et al. 2006). This change in nomenclature is
strongly supported by our study; for the remainder of this paper we refer to these galaxies as Polar Disk
Galaxies.


 The two scenarios for the formation of Polar Disk Galaxies which  have dominated the literature involve the interaction of two galaxies. One proposes that they form in a collisional merger between two galaxies \cite{bekki}. In the second scenario,  the polar disk forms from gas which is stripped from a donor galaxy which passes by the central galaxy, without merging \cite{schweizer}. Neither of the proposed galaxy interaction models, merger nor accretion from companions, has been shown to self-consistently explain the high gas mass within the polar structure, its extended nature, and the spatial coincidence of stars with this gas \cite{karataeva1}, with no evidence of interaction induced star bursts. The merger model is also difficult to reconcile with the presence of an inner disk as observed, for example, in NGC4650A \cite{iodice04} (disks are commonly believed to be destroyed by mergers). Further, both models predict that polar disk galaxies would reside preferentially in environments in which galaxy interactions are common, such as galaxy groups, yet such an environmental dependence is not observed \cite{brocca}. This is emphasized by the fact that, before being reclassified as a polar ring galaxy, NGC6822 was considered a typical isolated dwarf irregular \cite{demers}.  A third scenario of cold accretion claims that polar rings can be formed from the accretion of filamentary cold gas \cite{maccio}. A  study of a galaxy formed in cosmological galaxy formation simulations demonstrated that it is possible for  gas accretion  to be  perpendicular  to the central galaxy. Yet, as with models of accretion scenarios,  the simulated galaxy does not share  the detailed features of observed Polar Disk Galaxies, and the issue of the formation of these galaxies  ultimately remained unresolved.

As part of a project undertaken by our group at  the {\it N body Shop}  aimed at simulating a large sample of galaxies in a cosmological context, a simulated galaxy was found (serendipitously) having two concentric, almost perpendicular disks. This simulated galaxy, run within the ``concordant"  $\Lambda$-cold dark matter cosmology, shares the detailed features of observed Polar Disk Galaxies. When comparing these properties of our simulated Polar Disk Galaxies to those observed, it is important to mimic the  methods employed by observers as closely as possible. The techniques used in this study to artificially ``observe" our simulated  galaxy are state of the art.  Using the age and metallicity information of the star particles in the simulations, we determine how the light of our simulated galaxy would appear when observed  specific telescopes.  Using the age and metallicity information of the star particles in the simulations, we determine how the light of our simulated galaxy would appear when observed by the specific telescope used to observe the  real galaxy to which the comparison is made, including the important effects of dust.
We make particularly detailed comparisons with NGC4650A, considered the prototypical Polar Disk Galaxy \cite{whitmore}.

\section{Methods}
\subsection{Simulation code}

We have used the fully parallel, N-body, smoothed particle hydrodynamics (SPH) code GASOLINE \cite{wadsley} to compute the evolution of the collision-less and dissipative elements respectively. Here we outline its essential features, while the interested reader is referred to the literature for full details \cite{governato}.  GASOLINE computes gravitational forces using a treeÐcode that employs multi-pole expansions to approximate the gravitational acceleration on each particle \cite{barnes}. Time integration is carried out using the leapfrog method, which is a second-order symplectic integrator. In cosmological simulations, it is necessary to implement periodic boundary conditions, achieved in GASOLINE by employing a generalized Ewald {hernquist} method to fourth order.

SPH is a technique which uses particles to integrate fluid elements representing gas \cite{gingold,monaghan}. GASOLINE is fully Lagrangian, spatially and temporally adaptive and efficient for large numbers of simulation particles. The code includes radiative cooling and accounts for the effect of a uniform background radiation field on the ionization and excitation
state of the gas. The implemented cosmic ultraviolet background, following the Harrdt-Madau model \cite{haardt}, includes photoionizing and photoheating rates produced by QSOs, galaxies and POPIII stars starting at z = 10, consistent with the combination of the 3rd year WMAP results and the Gunn-Peterson effect in the spectra of distant quasars \cite{alvarez}. We use a standard cooling function for a primordial mixture of atomic hydrogen. The internal energy of the gas is integrated using the asymmetric formulation that gives results very close to those of a formalism that conserves entropy, but conserves energy better. Dissipation in shocks is modeled using the quadratic term of the Monaghan artificial viscosity \cite{wadsley}, with a Balsara \cite{balsara} correction term to reduce excessive artificial shear viscosity. 

A simple but physically motivated recipe \cite{stinson} describes star formation and the effects of subsequent energy feedback from supernovae. Adopted supernova and star formation efficiency parameters were previously tested, and we use a Kroupa Initial Mass Function \cite{kroupa}. Metal enrichment from both supernova types Ia \& II are followed based on yields from the literature \cite{raiteri}. The simulation described in this paper was run at the San Diego Supercomputing Facility using up to 512 CPUs, for a total of about 500k CPU hours. 

\begin{figure*}
\begin{center}
\includegraphics[width=6.3in]{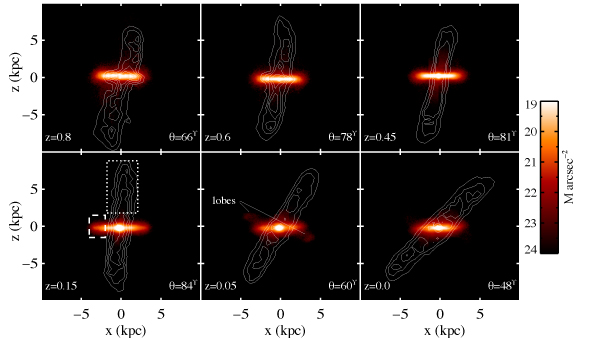}
\caption{At six time-steps during the polar disk evolution, we make edge-on surface brightness maps of the inner disk stars, i.e. that form between redshift $z=1.6$ and $z=0.7$, in the SDSS i-band filter. The color bar indicates the surface brightness level in Magnitudes/arcsec$^2$. We overplot a contour map of the cold gas ($<$4x10$^4$ Kelvin), where star formation takes place. At $z=0.8$, cold gas and hence ongoing star formation are present in the inner disk, but new cold gas accretion is highly inclined to this inner disk. The angle between this new forming polar disk and the inner disk is currently 66$^{\circ}$, as indicated in the bottom right corner. Between $z=0.6$ and $z=0.15$, the cold gas remains at an inclination of between 77-84$^\circ$ to the inner disk. The dashed box region at $z=0.15$ is selected for further analysis in Figure 2 as representative of the inner disk, whilst representative polar disk stars are within the dotted box. At $z=0.05$, a dynamically induced morphological feature becomes apparent in the inner disk, marked as ÒlobesÓ. By the present time, $z=0$, the polar and inner disk appear on their way toward aligning.
}
\label{evolve}
\end{center}
\end{figure*}

\subsection {Cosmological Initial Conditions}
The initial conditions for the galaxy described in this paper were obtained using  the so called volume renormalization technique to achieve higher resolution in a region of interest. It is a higher resolution realization of one the simulations already presented elsewhere \cite{brooks}. The virial mass of the halo that was selected at $z=0$ to be re-simulated at higher resolution is 1.6$\times$10$^{11}$M$_\odot$ (the virial mass is measured within the virial radius R$_{vir}$, the radius enclosing an overdensity of 100 times the critical density $\rho_{crit}$). The halo was originally selected within a low resolution dark matter only simulation run in a concordance, flat, $\Lambda$-dominated cosmology:  $\Omega_0 = 0.3$, $\lambda=0.7$, $h = 70$, $\sigma_8= 0.81$, where $\Omega_0$ is the matter density, $\Lambda$ the cosmological constant, $h$ is HubbleÕs constant, and  $\sigma_8$ is the root mean square fluctuations in the matter power spectrum at the 8Mpc scale. The size of the simulated region, 28.5 Mpc, is large enough to provide realistic torques. The power spectra to model the initial linear density field were calculated using the CMBFAST code to generate transfer functions. Dark matter, star and gas particle masses in the high-resolution regions are 9.4$\times$10$^{4}$, 3.3$\times$10$^{3}$ and 1.6$\times$10$^{4}$M$_\odot$, respectively. The force resolution, i.e. the gravitational softening, is 0.15 kpc. In total there are 1.4$\times$10$^{6}$ DM particles within the virial radius. At $z=0$ there are a total of 3.5$\times$10$^{6}$ baryonic particles within the central 20 kpc (each gas particle spawns up to three star particles and is then distributed between gas neighbors when its mass falls below 20\% its original one). Star particles lose mass due to supernovae and stellar winds. This mass is redistributed to nearby gas particles. With our choices of particle number and softening, the smallest subhalos resolved have typical circular velocities of 10\% of their host, ensuring that all subhalos able to retain even a small fraction of their baryons are resolved. The UV field makes ambient gas too hot to be bound to the lowest mass halos: most subhalos with mass below 5$\times$10$^{9}$M$_\odot$ are empty of stars. Particles in the high resolution region have gravitational spline softening evolved in a comoving manner from the starting redshift ($z =100$) until $z=9$, and kept then fixed from $z=9$ to the present. The softening values are a compromise between reducing two body relaxation and ensuring spatial resolution of disk scale lengths and the central part of dark matter halos. The simulation was run with an integration parameter $\eta=0.195$ and a treecode opening angle $\theta$ of $0.525$ for redshift $z>2$ and $0.725$ afterwards \cite{moore,power}. The simulated galaxy analyzed in this study is the highest resolution of its kind  in the published literature to date.

\subsection{Creation of Mock Images}
We used the open source software Sunrise \cite{jonsson} to generate our artificial optical images in Fig. 1, 3, 4, 7 \& 9. Sunrise allows us to measure the dust reprocessed spectral energy distribution (SED) of every resolution element of our simulated galaxies from the far UV to the far IR with a full 3D treatment of radiative transfer. Filters mimicking those on major telescopes such as the Hubble Space Telescope (HST)  are used to create mock observations. Sunrise uses Monte Carlo techniques to calculate radiation transfer through astronomical dust, including such effects in the determination of SEDs.

\section{Evolution of the Polar Disk}
Polar Disk Galaxies are only observed in the nearby Universe, due to inherent difficulties of observing these exotic galaxies, but they are presently in a variety of  evolutionary stages. This allows us to link properties of observed Polar Disk Galaxies to different phases in the simulation, and unravel a detailed history of their evolution. In the particular merging history of the simulation in which a polar disk forms, an inner disk starts forming shortly after the last major merger at redshift $z \sim 2$. The mass ratio of the merging galaxies is very close to 1:1 but, due to its gas rich nature, the galaxy rapidly forms a new disk whose angular momentum is largely determined by the merger orbital parameters. At later times gas continues to be accreted to the galaxy but in a plane that is almost perpendicular to the inner disk.

\begin{figure}
\begin{center}
\includegraphics[width=88mm]{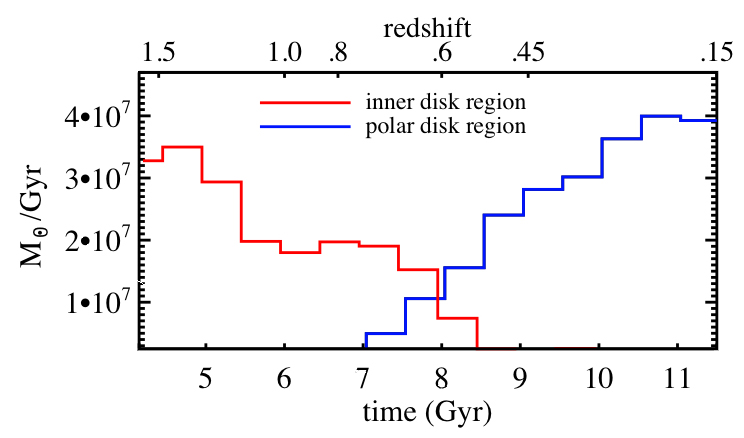}
\caption{Stars from a region of the inner disk at $z=0.15$  (dashed box, Fig. 1) and polar disk (dotted box, Fig. 1) are selected. We plot the star formation history within the two regions, since the time when the inner disk begins to form. Stars form reasonably steadily within the inner disk until $z\sim0.7$, and cease by $z\sim0.5$. This reflects the lack of cold gas in the inner disk region at later times, shown in Fig. 1. The polar disk stars began forming at $z\sim0.7$, with star formation relatively constant for 1.5 Gyrs between $z\sim0.5$ and $z\sim0.3$, and then constant at a slightly higher rate for a further 1.5 Gyrs between $z\sim0.3$ and $z\sim0.15$. }
\end{center}
\end{figure}

We first analyze a snapshot of the galaxy at redshift $z\sim 0.8$, shortly after the polar disk begins forming. At this time, the central galaxy is still forming stars in a disk, while the bulk of new star formation is in the highly inclined polar disk ($\sim 66^\circ$ to the inner disk, Fig. 1), with star formation continuing in both disks until $z=0.6$ (see Fig. 2). Two prominent spiral arms are evident in the polar structure. At this stage, the galaxy shares many features of NGC660, a local polar ring galaxy in which the central galaxy is still forming stars. The polar structure in both NGC660 and the simulation is a disk with an exponential light profile rather than a ring, and has a young stellar population with continuous star formation \cite{vandriel}.  By $z=0.5$, the inner disk has exhausted its gas, whilst gas continues to fall onto the polar disk. From this point in time to the present, the galaxyÕs star formation occurs almost exclusively in the polar disk (Fig. 2). The polar disk continues to accrete material, at an inclination which remains stable between 78-84$^\circ$ until  $z=0.15$. Almost all of the neutral hydrogen, as seen in Fig. 1, is perpendicular to the inner disk during this time. The likeness to the HI maps of UGC9796 \cite{cox06}, and UGC7575 \cite{sparke},  when also over-plotted  on optical images which highlight their inner disk, is striking (see also section~\ref{hydrogen}).

The polar disk in our simulation is stable for at least the 3 Gyrs between $z=0.6$ and $z=0.15$. This is well supported by observations of stellar populations in several polar disks, which indicate that such structures are long lived. For example, the stellar population of the polar disk in NGC4650A has most likely been continuously forming for stars 3 Grys \cite{gallagher,mould}, and the star formation observed in the polar disk of UGC9796 indicates that it is 3-5 Gyrs old \cite{cox06}. Fig. 3 shows the stunning correspondence between the simulated Polar Disk Galaxy at $z=0.15$ and NGC4650A. At this time, the simulated Polar Disk Galaxy and NGC4650A both have extended polar disk structures with young stellar populations, ongoing star formation, blue colors, spiral arms, $M(HI)/L_B$ ratios that are  typical of low surface brightness galaxies \cite{iodice02}, and almost flat rotation curves. Both have rotationally supported S0 like inner galaxies, which have had little or no star formation over a period of 3 Gyrs, and exponential light profiles \cite{iodice04}. Details of these properties  of the inner and polar disks of the simulation are presented in   section~\ref{details}.

\begin{figure}
\begin{center}
\includegraphics[width=85mm]{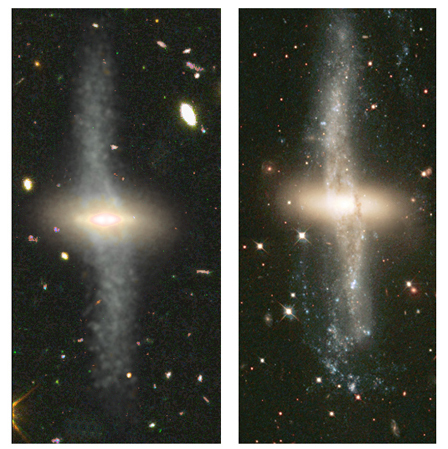}
\caption{The simulated polar disk (left) is imaged by assigning three mock HST bands, at 450, 606, and 814 nm, the colors blue, green and red, respectively. The image is superimposed onto an HST background. NGC4650A, considered the prototypical polar disk galaxy, is imaged (right) using the same color-band assignment \cite{gallagher}. The red color of the inner disks reflects the old age of their stars, which emit longer wavelength light. The younger stars of the polar disks are more prominent in the shorter wavelength bands, hence their blue hue.}
\end{center}
\end{figure}

\begin{table*}
\begin{center}
\begin{tabular}{cclcccccc}
\hline
\hline
& I$_{AB}^{a}$ & B-I$^{b}$  & Vrot$^{c}$ & V/$\sigma_V$$^{d}$&h$_l$$^e$  & Stellar Mass$^{f}$& HI Mass & M(HI)/L$_B^{g}$\\
${}$ &   &   & km/s &   & kpc & M$_\odot$ & M$_\odot$ & M$_\odot$/L$_\odot$ \\
\hline
Galaxy& -20.3 &  1.7    &   -        &  -         & -      & 1.0 $\times$ 10$^{10}$& 2.1 $\times$ 10$^{9}$  &  0.80 \\
Inner &    -19.8  &  1.6    &   105  & 4.4     & 0.95   & 3.5 $\times$ 10$^{9}$ &-&-  \\
Polar &       -19.2      &  1.3    &   115  & 1.4     & 3.4   &  1.6 $\times$ 10$^{9}$ & 2.1 $\times$ 10$^{9}$ & 1.4 \\
\hline
\end{tabular}
\end{center}
\caption\footnotesize {{Summary of the properties of the simulated galaxy,its central S0 galaxy and the polar disk at $ z=0.15$, when the simulation most resembles NGC4250A.  (a) inner and polar absolute magnitudes are derived by calculating contributions to flux of stars born before and after $z=0.8$, after which star formation is mainly in the polar disk. (b) Inner and polar disk colors are for the regions indicated at $z=0.15$ in Fig. 1 (c) Line of sight velocity at 2.2 scalelengths (see Fig. \ref{rot})  , (d)  at 2.2 scalelengths, (e)  see Fig. \ref{prof} (f) a significant amount of galaxy stellar mass is in an old, faint bulge/spheroid, (g)  for the region of the polar disk indicated at $z=0.15$ in Fig.~\ref{prof}}
}
\label{parameters}
\end{table*}

During the final 1.5 Gyrs to the present, our simulated polar disk becomes less stable, and in particular the inner regions of the polar disk begin to dynamically interact with stars in the central galaxy. This creates a complex dynamical system with the polar and inner structures becoming less distinct.  The inner regions of the Polar Disk Galaxy ESO-603-G21 show similarly complex and overlapping dynamical structures \cite{reshetnikov02}. At $z\sim0.05$, the inner disk of our simulated galaxy forms a morphological feature remarkably similar to those described as ``lobes" in the central galaxy of ESO-603-G21 (Fig. 1). These are old stars that have been dynamically stirred from the inner disk by the interaction of the two disks during this stage of the galaxyÕs evolution. Interestingly, ESO-603-G21 also has a polar disk that is prominent in the K-band \cite{reshetnikov02}, indicating that, like the polar disk component of our simulation at this time, it also is several Gyrs old.

\section{Properties of the Inner and Polar Disks}\label{details}
We refer to various properties of  the two orthogonal components of our simulated polar disk galaxy in the previous section, in which we follow the galaxy's evolution. We show plots which verify these properties in this section, presented in a manner which closely mimics observations, facilitating direct comparisons with observed polar ring galaxies. In several plots, we provide distances in kiloparsecs as well as arcseconds. Arcsecond scales are derived by assuming that the simulated galaxy has the same heliocentric systemic velocity  as observed for NGC4650A \cite{arnoboldi97}, and using  70 km/s/Mpc for the Hubble constant, which implies that 1$''$= 201pc. Table 1 provides a summary of the properties of the simulated galaxy, its inner (S0 type) disk  and the polar disk at $z=0.15$, when the simulation most resembles NGC4250A


\begin{figure}
\includegraphics[width=75mm]{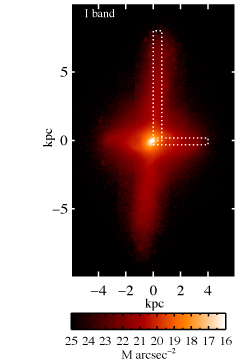} 
\caption{ The two perpendicular regions over which surface brightness profiles are taken in Fig.~\ref{prof} are indicated by the rectangles in this I-band surface brightness map. These regions are also used for deriving line of sight rotation curves in Fig.~\ref{rot}.  }
\label{region}
\end{figure}

\begin{figure}
\begin{center}
\includegraphics[width=70mm]{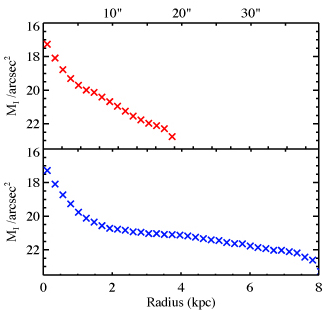}
\end{center}
\caption{ Radial surface brightness profiles in the I-band are plotted for the central galaxy (red crosses, upper panel) and polar disk (blue, lower panel).  The two perpendicular regions over which the profiles are taken are indicated by the rectangles in Fig.~\ref{region}.  }
\label{prof}
\end{figure}

\begin{figure}
\begin{center}
\includegraphics[width=73mm]{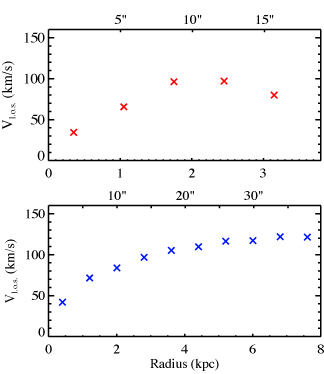}
\end{center}
\caption{ Line of sight velocity, versus radius, in the two regions indicated in Fig.~\ref{prof}.  Radial distances are provided in both kpcs and arcseconds. The inner disk velocities (red crosses, upper panel) are derived using stars, while the polar disk velocities (blue crosses, lower panel) were derived using cold gas (T$<$40,000K) .  }
\label{rot}
\end{figure}

\subsection{Light Profiles}
The radial surface brightness profiles of both components of our simulated polar disk galaxy attest to their disk galaxy nature. We plot  these profiles, in Fig.~\ref{prof}, in two perpendicular regions as indicated in the left panel, at $z=0.15$.   Both components have exponential light profiles outside the central bulge, with scalelengths of  the inner and polar disks of approximately 0.95 and 3.4 kpc, respectively. In order to facilitate easy comparison with observations \cite{iodice02}, we assume that the galaxy is at the same distance as NGC4650A. The radius in arcseconds indicated in the upper axis of the plot.

\subsection{Rotation Curves}
We derive rotation curves of the inner and polar disk galaxies by using line of sight velocities of the stars and cold gas (T$<$40,000K) respectively, in the regions indicated in Fig.~\ref{prof}. Both the inner disk and polar disk rotation curves become reasonably flat. The rotation velocity at 2.2 scalelengths is very similar for both components, $\sim$105 km/s and $\sim$ 115 km/s for the inner and polar disks respectively. Radial distances are provided in kpc as well as arcseconds to facilitate comparison with observations \cite{swaters,iodice06} of NGC4650A. Both components are rotationally supported; $V/\sigma_{V}$ at 2.2 scalelengths are 1.3 and 4.4 for the inner and polar disks respectively.

\subsection{HI warps and position-velocity  maps}\label{hydrogen}

\begin{figure}
\includegraphics[width=68mm]{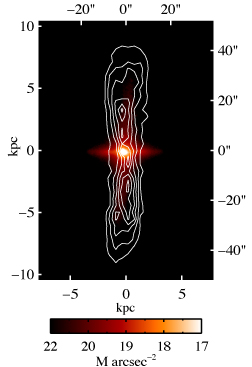}
\caption{ Surface brightness map of the simulation  in the SDSS r-band with the HI density map overplotted, shown at $z=0.15$.}
\label{hi}
\end{figure}

\begin{figure}
\includegraphics[width=68mm]{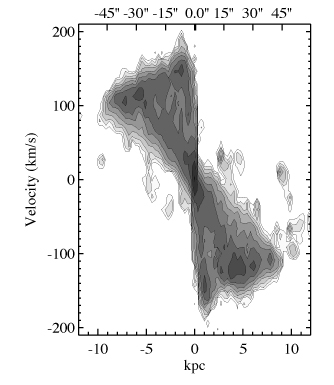}
\caption{ HI position-velocity map of the simulation at the same epoch. }
\label{posvel}
\end{figure}

The neutral Hydrogen (HI) maps of the simulations are plotted in Figs.~\ref{hi} \& \ref{posvel}. Fig.~\ref{hi} is an  SDSS r-band surface brightness image, which highlights the old, central galaxy stars, with contour lines of HI gas overlaid. Almost all the HI is in the polar disk, virtually  perpendicular to the central galaxy. The simulation is  shown here at z=0.15, and presents greater detail than seen in Fig.~1. The morphological features of the HI maps are very similar to those observed in  polar disk galaxies such as UGC9796 \cite{cox06} and UGC7575 \cite{sparke}.  An integral shaped warp is apparent in these HI maps of the simulated and observed polar disk galaxies. Such warps are evident at various times in the simulation.

The HI position-velocity map for  simulation at $z=15$ is shown in Fig.~\ref{posvel}, computed by projecting the data onto the major axis of the polar disk.  Our simulated galaxy shares many qualitative features as the position-velocity map of  NGC 4650A \cite{sparke}, in particular the characteristics of a disk rather than ring: modeling an infinitely narrow ring would produce a linear position-velocity diagram.

\subsection{Spiral Arms}
\begin{figure*}
\begin{center}
\includegraphics[width=160mm]{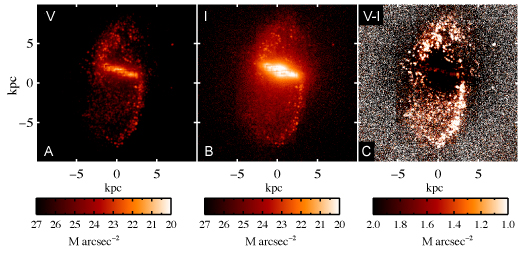}
\end{center}
\caption{ The V-band image (panel a) is sensitive to recent star formation, and shows the central galaxy's inner disk almost edge on.  Two prominent spiral arms are evident in the polar structure.  The longer wavelength light imaged in the I-band (panel b) is more sensitive to older stellar populations. The V-I image (panel c) is derived by subtracting I band image from the V-band, and  highlights recent star formation. 
\\
}
\label{spiral}
\end{figure*}

 The polar disk of the simulated polar disk galaxy has its most prominent grand design spiral arms at early times in its evolution. We highlight the two spiral arms as they exist at $z=0.6$ in the simulation, in Fig.~\ref{spiral}. The surface brightness in the V-band image (panel A) is sensitive to recent star formation, and shows the polar disk at an angle of 40$^\circ$  to edge on, and the central galaxy's inner disk almost edge on.  The longer wavelength light imaged in the I-band (panel B) is more sensitive to older stellar populations. By subtracting the I-band image from the V-band, the difference, V-I image (panel C) highlights recent star formation, which occurs largely in the spiral arms. It is interesting that NGC4650A also has two grand design spiral arms \cite{arnoboldi97}. 

\section{Discussion}

The misalignments of angular momentum vectors between early and later accreted material is expected to be common in Cold Dark Matter structure formation, but in general the misalignment is small \cite{quinn}. In all our disk galaxy simulations, we see smaller misalignment of the inner disk, and later forming outer disk, at some time during its formation, resulting in a transient ``integral sign" warp in the galaxy. Such warps are seen in over half of observed disk galaxies \cite{bosma}, indicating that these processes of misalignment witnessed in our simulations are real.  Polar disks are the rare cases of such processes where the inclination is extreme. Such polar inclinations are seen to be stable, as is the case in our simulation 

It is telling that Polar Disk Galaxy ESO-325-G58, which is observed face on,  was classified as a barred spiral before it was shown that the ``bar" was in fact an inner disk perpendicular to the line of sight \cite{iodice04}.  Similarly, NGC6822 is a local galaxy that was long classified as a barred irregular galaxy, but has recently been shown to be a polar ring galaxy \cite{demers}. An interesting observational challenge suggested by our study is to determine the number of face on Polar Disk Galaxies that have been classified as barred galaxies. 

This scenario also has important implications for studying the nature of dark matter; Polar Disk Galaxies are used as a probe of the shape of dark matter halos \cite{casertano,schweizer,sackett,iodice06}. Our study implies that we now have the important added ability to determine how the shape of the dark halo is oriented with respect to the direction of cosmological gas flow, and how these relate to the surrounding distribution of matter in which a galaxy is embedded. 

The processes of galaxy merging and accretion are both features of a Cold Dark Matter dominated  universe and both are capable, indeed necessary, in producing observed  properties of galaxies. The  galaxies we have referred to in this study have been characterized by polar structures with disk galaxy characteristics, and we have shown that such galaxies form from cosmological gas infall. In other galaxies, such as  AM 2020-504 \cite{whitmore,arnaboldi93},  the polar structure is better described as a ring, with gas and stars in a narrow annulus. Deciding which galaxies are classified as polar rings and which as polar disks is an observational problem, which is best accomplished by comparing the polar structure to the outer regions of late type galaxies. Do Polar Ring and Polar Disk Galaxies form in the same manner?  There are two ways in which narrow polar structures may form within our model for Polar Disk Galaxies. The passage of a small, satellite galaxy through a galactic  disk is capable of transforming the disk into a ring.  The spectacular ``Cartwheel" galaxy is a proto-type for Ring Galaxies  \cite{zwicky}. This mechanism is capable of forming a Polar Ring Galaxy from a Polar Disk Galaxy. It is also possible that gas infall along a filament may have a narrow angular momentum distribution, and naturally settle into a ring. But  our study is not conclusive on this issue, and the classic merging and accretion models remain viable explanations for the formation of  narrow polar ring structures. For now, it may be necessary to treat Polar Rings and Polar Disks  as two distinct galaxy types. 

Most of the comprehensively studied polar structures are disks, and the evidence linking the polar disk structures to normal disks is compelling.  The irresistible conclusion of our simulation is that the formation processes of Polar Disk Galaxies are the same as for the outer disk regions of normal disk galaxies, except the polar disk forms in a different plane to the central galaxy. Our scenario naturally explains the properties of several extensively studied Polar Disk Galaxies in terms of the different stages of their evolution. The notion of  ``rebuilding a disk" is an essential ingredient in galaxy formation models, and has been previously demonstrated in simulations \cite{steinmetz}. Our study implies that Polar Disk Galaxies are a spectacular demonstration of the rebuilding of disks through accretion of gas from the ``cosmic web" predicted by hierarchical models of galaxy formation

\acknowledgements We thank Enrica Iodice for helpful discussions, and Brad Gibson and Julianne Dalcanton  for feedback on a draft.  C.B., F.G., T.Q. \& A.B. are supported by an NSF ITR grant, PHY-0205413. F.G. was supported by NSF grant AST-0607819 and by a Spitzer Theory Grant. The simulation was run at the San Diego Supercomputing Facility.  

\fontsize{10}{10pt}\selectfont

\end{document}